\documentclass[aps,preprint,superscriptaddress,groupedaddress]{revtex4-2}
\usepackage[utf8]{inputenc}
\usepackage{tabularx}
\usepackage{array}
\usepackage{amsmath}
\usepackage{graphicx}
\graphicspath{{pictures/}}
\usepackage{xfrac}
\usepackage{ulem}
\usepackage{float}
\usepackage{wrapfig}
\usepackage{caption}
\usepackage{subcaption}
\usepackage[dvipsnames]{xcolor}
\usepackage{mathrsfs,amsmath,amssymb,bm}
\allowdisplaybreaks
\usepackage{comment}

\begin{document}
\title{Connecting finite-time Lyapunov exponents with supersaturation and droplet dynamics in a turbulent bulk flow}
\author{Vladyslav Pushenko and J\"org Schumacher}
 \affiliation{Institut für Thermo- und Fluiddynamik, Technische Universität Ilmenau, Postfach 100565, D-98684 Ilmenau, Germany}
 \date{\today}
 
\begin{abstract}
    The impact of turbulent mixing on an ensemble of initially monodisperse water droplets is studied in a turbulent bulk which serves as a simplified setup for the interior of a turbulent ice-free cloud. A mixing model was implemented which summarizes the balance equations of water vapor mixing ratio and temperature to an effective advection-diffusion equation for the supersaturation field $s({\bm x},t)$. Our three-dimensional direct numerical simulations connect the velocity and scalar supersaturation fields in the Eulerian frame of reference to an ensemble of cloud droplet in the Lagrangian frame of reference. The droplets are modeled as point particles with and without effects due to inertia. The droplet radius is subject to growth by vapor diffusion. We report the dependence of the droplet size distribution on box size, initial droplet radius, strength of the updraft, with and without gravitational settling. In addition, the three finite-time Lyapunov exponents $\lambda_1\ge\lambda_2\ge \lambda_3$ are monitored which probe the local stretching properties along the particle tracks. In this way, we can relate regions of higher compressive strain to those of high local supersaturation amplitudes. For the present parameter range, the mixing process in terms of the droplet evaporation is always homogeneous, while it is inhomogeneous with respect to the relaxation of the supersaturation field. The probability density function of third finite-time Lyapunov exponent, $\lambda_3<0$, is related to the one of the supersaturation $s$ by a simple one-dimensional aggregation model. The probability density function (PDF) of $\lambda_3$ and droplet radius $r$ are found to be Gaussian while the PDF of the supersaturation field shows sub-Gaussian tails.
\end{abstract}
\maketitle

\section{Introduction}
Atmospheric clouds are complex physical systems which play a key role on synoptic time scales for the formation of precipitation \cite{Shaw2003} and on much longer time scales for the radiation budget of the climate of our planet \cite{norris2016evidence}. The water appears in different phases inside a cloud,  either as water vapor, liquid water droplets, or  ice crystals. All phases affect the optical properties of clouds crucially; the overall turbulent dynamics determines the local in- and decrease of the corresponding mixing ratios \cite{kokhanovsky2004optical,grabowski2013growth}. Warm ice-free clouds form a two-component turbulent multiphase flow (water and vapor). The cloud water droplet number density and size distribution are then  two key quantities in view to rain formation for which droplets typically have to grow up to about 1 mm. One of the central questions remains how the turbulent mixing of dry air and moist air, in particular entrainment \cite{Blyth1993,Mellado2017}, alters the number density and size distribution of cloud water droplets inside the cloud and which role the locally fluctuating supersaturation does play, see e.g. refs. \cite{andrejczuk2006numerical,Celani2008,Lanotte2009,sardina2015continuous,Gotoh2016,grabowski2017broadening,Saito2018,Fossa2022,Grabowski2022} for numerical investigations inside clouds or at its edges.

These questions provide the central motivation for the present study. Here, we discuss a strongly simplified model of turbulent mixing in the bulk of a cloud by means of three-dimensional direct numerical simulations (DNS). This simplification consists of a summary of the balance equations for the Eulerian fields of vapor mixing ratio and temperature, $q_v({\bm x},t)$ and $T({\bm x},t)$,  to an effective balance equation for the scalar supersaturation field, following refs. \cite{celani2005droplet,Lanotte2009,sardina2015continuous,fries2021key}. We assume that the supersaturation field $s({\bm x},t)$ does not act back on the flow; it is thus a passive scalar field in the present approach. The  field is given by 
\begin{equation}
s({\bm x},t)= \frac{q_v({\bm x},t)}{q_{vs}(T)}-1\,,
\end{equation}
with the saturation mixing ratio $q_{vs}$ of vapor which depends on temperature $T$ via the Clausius-Clapeyron equation \cite{yau1996short}. This field determines the diffusion growth of an ensemble of individual Lagrangian cloud water droplets which are considered as point-like particles with an attached radius coordinate. In the present work, we connect the statistics of finite-time Lyapunov exponents of the advecting velocity field \cite{pikovsky2016lyapunov}, which quantify the local stretching and compression rates in the Lagrangian frame of reference, to the statistics of $s({\bm x},t)$ and thus to the droplet size (or droplet radius) distribution. A further simplification of the present model is that the droplets are subject to a one-way coupling only. \citet{Goetzfried2017} showed that the particle Reynolds number for the initial radii, that are taken here, remain smaller one. An objective of our approach is to test an aggregation model of the turbulent mixing of passive scalar fields which has been successfully applied for turbulent mixing at high Schmidt numbers ${\rm Sc}$ in the past \cite{Villermaux2003,duplat2008mixing,gotzfried2019comparison}. The scalar statistics follows then by a successive stretch-twist-fold stirring of scalar concentration filaments that is subject to molecular diffusion.  We report current results for a small cubical reference volume in the bulk of a cloud where a Schmidt number ${\rm Sc}\approx 0.7$ is considered for the scalar supersaturation field $s({\bm x},t)$. As already said, this is significantly smaller than the ones in refs. \cite{Villermaux2003,duplat2008mixing,Meunier2010,Meunier2022}. We find that the supersaturation dynamics is then Gaussian-distributed in the core with sub-Gaussian tails, both, from the Eulerian statistical analysis as well as along the individual droplet trajectories. This causes Gaussian-shaped droplet size distributions which broaden faster when the box size becomes bigger. 
 
Turbulent mixing, particularly across the boundary of a cloud, influences the number density and size distribution of cloud droplets by causing fluctuations of water vapor and liquid water content, as we showed in previous works \cite{Kumar2013,kumar2012extreme,Kumar2014,Goetzfried2017,kumar2018scale}, partly in significantly more complex configurations with several nonlinear feedbacks. Here, we focus on the simpler mixing inside the cloud bulk. Mixing can be characterised by Damköhler number $\textrm{Da}$ \cite{siewert2017statistical,pinsky2016theoretical}, the ratio of the typical flow time scale, such as the large-scale eddy turnover time to a characteristic thermodynamic reaction time scale of interest,
\begin{equation}
{\rm Da}=\frac{\tau_L}{\tau_{\rm react}}\,.
\end{equation}
Thermodynamic processes, which are characterised by small Damk\"ohler numbers $({\rm Da}\ll 1)$, proceed slower than the flow is mixed; this mixing regime is called thermodynamics-limited or {\em homogeneous}, see e.g. ref. \cite{Burnet2007}. For cloud droplet evaporation, it means that moist air will be properly mixed and droplets will evaporate at approximately the same rate. In the regime with the large Damk\"ohler number ($\textrm{Da}\gg1$), the corresponding thermodynamic regime is called  mixing-limited  or {\em inhomogeneous}. In this regime, the moist air is mixed slower by the flow than it is saturated by evaporating droplets. For inhomogeneous mixing, cloud droplets evaporate in different regions with different rates \cite{jensen1989simple,kumar2012extreme,Beals2015,fries2021key}. As it was shown in \cite{kumar2012extreme}, the inhomogeneous mixing gives extended tails of the cloud droplet size distribution.  Higher levels of fluctuations of water vapor content in different regions generate different rates of growth and shrinkage of cloud water droplets. We want to investigate how strong this variability of the droplet radius $r$ is in the bulk of a cloud.

The paper is organized as follows. In Section II, we provide the cloud mixing model, define the main parameters of our model and discuss their influence on the mixing processes. Section III discusses DNS results obtained from the turbulence fields in the Eulerian frame of reference. Section IV is dedicated to the Lagrangian analysis of the cloud water droplets. While subsection IV A discusses the Lagrangian tracer case, subsection IV B investigates effects of gravity and initial radius on the droplet dynamics. The definition of finite-time Lyapunov exponents (FTLE) is provided in Sec. V. Here, we connect these results to those of the previous sections before a summary and outlook is given in Sec. VI. The appendix provides the derivation of the effective advection-diffusion equation for the supersaturation field for completeness.

\section{Numerical simulation model}

\subsection{Eulerian and Lagrangian model equations}
In our cloud mixing model, we consider a subvolume $V=L^3$ in the bulk of a cloud as a multiphase system which consists of dry air, water vapor and liquid water. We assume periodic boundary conditions in all three directions. The turbulent velocity field ${\bm u}({\bm x},t)$ is assumed to be statistically stationary, homogeneous, and isotropic. The full complexity of the mixing process in presence of phase changes requires balance equations for the (1) temperature field $T({\bm x},t)$ including latent heat release, for the (2) vapor mixing ratio $q_v({\bm x},t)$ including condensation rate as a loss term, and (3) liquid water mixing ratio $q_l({\bm x},t)$ including condensation rate as a gain term. 

The models of \citet{celani2005droplet,Celani2008}, \citet{sardina2015continuous} and \citet{fries2021key} simplify this complex dynamics in three aspects. First, the liquid water content is represented by an ensemble of individual spherical point-like cloud water droplets. Attached is a droplet radius which can increase and decrease, thus changing $q_l$. Secondly, the fields $T$ and $q_v$ are summarized in the scalar supersaturation field $s({\bm x},t)$. Thirdly, the mixing of the scalar field and the advection of the droplets will not couple back to the turbulent velocity field ${\bm u}({\bm x},t)$ via a buoyancy term, i.e., there is a one-way coupling considered only. This assumption is justified when the droplet radii remain small and the droplet number density $n\sim 100$ cm$^{-3}$ as in our study. The Eulerian equations of motion follow to
\begin{subequations}
        \label{eq:final_eq_eulerian}
        \begin{align}
            \label{eq:incompress}
            {\bm \nabla} \cdot {\bm u} &= 0,\\
            \label{eq:ns_final}
            \frac{\partial{\bm u}}{\partial t}+({\bm u}\cdot{\bm \nabla}){\bm u} &= - \frac{{\bm \nabla} p}{\rho} + \nu {\bm \nabla}^2 {\bm u} + {\bm f},\\
            \label{eq:s_full}
            \frac{\partial s}{\partial t}+ ({\bm u}\cdot{\bm \nabla})s &= D_{s} {\bm \nabla}^2 s + A_1 u_z - A_2 \frac{4 \pi \rho_{L} K'}{V_{a}} \sum_{i=1}^N  r_i(t) s({\bm X}_{i},t).
        \end{align}
\end{subequations}
\begin{table*}
    \caption{The parameters of the simulations with different domain sizes $L$ and energy injection rates $\epsilon_{\rm in}$, numerical resolution as a ratio of bin size $\Delta x$ to obtained Kolmogorov length scale $\eta_K$, number of grid-points in each direction $n$, umber of particles $n_p$, coefficient of vertical ascending $A_1$, time- and volume-averaged  root-mean square velocity magnitudes $U$, large-scale eddy turnover times $\tau_L$, and time- and volume-averaged root-mean square supersaturation values $s_{rms}$. Furthermore, listed are Reynolds numbers $\textrm{Re}$, Taylor microscale Reynolds numbers $R_\lambda$, and the Damköhler numbers, $\textrm{Da}_\textrm{s}$ and $\textrm{Da}_\textrm{d}$, specified for supersaturation and droplet evaporation, respectively. Quantities are given in their physical dimensions with amplitudes that correspond to typical conditions in a cloud.}
    \begin{ruledtabular}
        \begin{tabular}{lccccccccccccc}
            Run & $L$& $\epsilon_{\rm in}$& $\dfrac{\Delta x} {\eta_K}$  & $n$ & $n_p$ & $A_1$ &  $U$ & $\tau_L$ & $s_{\rm rms}$ & Re & $R_\lambda$ & $\textrm{Da}_s$ & $\textrm{Da}_d$ \\
            & $[m]$& $[m^2/sec^{-3}]$&  & & & $[m^{-1}]$ & $[m/sec]$ & $[sec]$ &   &   &  &  &\\ \hline
            1& 0.128 & 0.0034 & 1.009 & 128 & 209,715 & 0.2 &  0.072 & 1.78 & 0.00439 & 614  & 28.4  & 1.04  & 0.0043\\
            2& 0.256 & 0.0034& 1.006 & 256 & 1,677,721 & 0.2 & 0.097 & 2.64 & 0.00795 & 1655 &  51.6  & 1.54  &  0.0125\\
            3& 0.512 & 0.0034& 1.002  & 512 & 13,421,772 & 0.2 & 0.129 & 3.97 & 0.01088 & 4406 & 72.9 & 1.85 & 0.5952\\
        \end{tabular}
    \end{ruledtabular}
    \label{table:parameter_study}
\end{table*}
Due to its small size, cloud water droplets will have a small Stokes number ($\textrm{St}\ll 1$) and are thus approximated as Lagrangian tracer particles without inertia following perfectly the streamlines of the turbulent velocity field for most of the work, except in subsection IV B. In ref. \cite{Kumar2013}, we showed that effects of additional droplet inertia for the presently chosen parameters remain small. The dynamics of the $N_0$ individual cloud water droplets is given by \cite{yau1996short}
\begin{subequations}
        \begin{align}
            \label{eq:lagr_tracer}
            \frac{d {\bm X}_i}{dt} &= {\bm u}({\bm X}_{i},t), \qquad i = 1,\dots,N_{0}\\
            \label{eq:radius}
             r_{i}\frac{dr_{i}}{dt} &= K' s({\bm X}_{i},t).
        \end{align}
        \label{eq:dimensional_eq}
\end{subequations}
The coefficients in the combined Euler-Lagrangian model equations depend on thermodynamic properties of the cloud. They are defined as
\begin{subequations}
    \label{}
    \begin{align}
             A_1 &= \frac{{\cal L} g}{R_{v} c_{p} T^2},\\
             A_2 &= \frac{R' T}{\varepsilon e_{s}(T)} + \frac{{\cal L}^2 \varepsilon}{p T c_{p}},\\
             K' &= \left[ \frac{{\cal L} \rho_{l}}{kT}\left(\frac{\cal L}{R_{v} T} - 1\right) + \frac{\rho_{l}R_{v}T}{D e_{s}(T)} \right]^{-1}.
    \end{align}
\end{subequations}
Here $p({\bm x}_i,t)$ the pressure field, $\rho$ the constant fluid density. Quantities $r_i(t)$ and ${\bm X}_i(t)$ are the radius and the spatial position for $i$-th cloud water droplet, respectively. Furthermore, $\nu$ is the kinematic viscosity, ${\bm f}({\bm x},t)$ the large-scale volume forcing of the flow, $D_{s}$ the diffusion coefficient of the supersaturation field, ${\cal L}$ the latent heat of evaporation, $g$ the acceleration due to gravity, $R_v$ the gas constant for water vapour, $c_p$ the specific heat at constant pressure, $R'$ the gas constant of dry air, $\varepsilon=R'/R_v$ the ratio of both gas constants, $e_s(T)$ the saturation water vapor pressure, and $\rho_l$ the density of liquid water. Also, $k$ is the thermal conductivity of air, $D$ the diffusion coefficient of the water vapor diffusion, $V_a$ the grid cell volume, and $N$ the number of particles in the vicinity of grid cell volume around position ${\bm x}$. 

The diffusion coefficient of the supersaturation field, $D_s$, can be well approximated by the diffusion coefficient of water vapour, $D$, \cite{sardina2015continuous}. The first of the two forcing terms, $A_1 u_z$, in eq. \eqref{eq:s_full} is due to the temperature gradient in $z$-direction. For the present box sizes, the prefactor turns out to be very small with ${\cal O}(10^{-3})$. Thus we have enhanced this factor in the first three runs and discuss the realistic magnitude in subsection IV B. The second term in eq. \eqref{eq:s_full} is a condensation rate term, i.e., a change of the liquid water content in the air slab. It quantifies the effects of condensation and evaporation and thus couples the dynamics of the scalar field and the cloud droplets. In turn, their individual radius changes with variations of the supersaturation field \cite{yau1996short}. In the appendix, we provide the detailed derivation of eq.~\eqref{eq:s_full} for completeness, see also \cite{sardina2015continuous,Chandrakar2020,fries2021key}.

\subsection{Dimensionless form of equations and parameters} 
We set the same initial radius $r_0$ for all droplets and seed them randomly across the whole computational domain. In order to get the dimensionless equations out of \eqref{eq:dimensional_eq}, the following characteristic scales are chosen. The root mean square of the velocity field is $U=\sqrt{2\,E_k/3}$, where $E_k= \frac{1}{2} \langle u_{i}^2 \rangle_{V,t}$ is the turbulent kinetic energy. As the characteristic time scale the large-eddy turnover time, $\tau_L=E_k\epsilon$, is chosen. Here $\epsilon = 2 \nu S_{ij} S_{ij}$ is a turbulent energy dissipation rate, where $S_{ij} = (\partial_j u_i +\partial_i u_j)/2$ is the rate-of-strain tensor. The characteristic passive scalar scale is the root mean square of the supersaturation, $s_{\rm rms}=\langle s^2\rangle_{V,t}$. Note that the supersaturation field is kept in a statistically stationary state. Here, $\langle\cdot\rangle_{V,t}$ is a combined average with respect to volume and time. The initial droplet radius is $r_0$, the corresponding number density is $n_0$. Using these assumptions, the dimensionless versions of \eqref{eq:final_eq_eulerian} and \eqref{eq:dimensional_eq} are given by
 \begin{subequations}
        \begin{align}
           {\bm \nabla} \cdot {\bm u} &= 0,\\
           \frac{\partial {\bm u}}{\partial t} +({\bm u}\cdot {\bm \nabla}){\bm u} &= -{\bm \nabla} p + \frac{1}{\rm Re} {\bm \nabla}^2 {\bm u} + {\bm f}\,,\\
            \frac{\partial s}{\partial t }+({\bm u}\cdot {\bm \nabla}) s  &=  \frac{1}{\rm Re\, Sc}  \nabla^2 s + \tilde{A}_1 u_z + \textrm{Da}_s V\sum_{i=1}^N r_i(t) s({\bm X}_i,t)\,,\\
            \label{eq:lagr_part_tr}
            \frac{d {\bm X}_i}{dt} &= {\bm u}({\bm X}_{i},t), \qquad i = 1,\dots,N_{0}\,,\\
            r_{i}\frac{dr_{i}}{dt} &= \frac{\textrm{Da}_d}{2} s({\bm X}_{i},t)\,,
        \end{align}
        \label{eq:non_dim_eq}
    \end{subequations}
with $\tilde{A}_1=A_1 U/\tau_L$. From (\ref{eq:non_dim_eq}), the flow depends on the following non-dimensional parameters, the large-scale Reynolds number Re, the Schmidt number Sc and both Damköhler numbers which are given by
\begin{equation}
        \label{eq:param}
        \textrm{Re} = \frac{U^2\tau_L}{\nu},\quad
        \textrm{Sc} = \frac{\nu}{D_{s}},\quad
        \textrm{Da}_s = \frac{\tau_L}{\tau_s},\quad
        \textrm{Da}_d = \frac{\tau_L}{\tau_d}\,.  
\end{equation}
These two Damk\"ohler numbers have been identified in refs. \cite{siewert2017statistical,pinsky2016theoretical}. Here, $\tau_\textrm{s}$ is a supersaturation relaxation time, the time scale at which supersaturation will decay to saturation state; $\tau_\textrm{d}$ is droplet evaporation time, which describes a timescale at which a droplet with initial radius $r_0$ should evaporate in a subsaturated environment with supersaturation magnitude $s=s_{rms}$. The two time scales are determined by
\begin{subequations}
    \begin{align}
        \tau_s &= \frac{1}{A_2 4 \pi \rho_{L} K' r_0 n_0},\\
        \tau_d &= \frac{r_0^2}{2 K' s_{\rm rms}}.
    \end{align}
\end{subequations}

When $\textrm{Da}_d \ll 1$, the cloud droplets evaporate at approximately the same rate because the flow is well-mixed. The other regime with $\textrm{Da}_d \gg 1$ implies that the droplets will have different rates of evaporation.

\section{Statistics and structure of the Eulerian turbulence}
 The Eulerian equations of motion are solved by a standard pseudospectral direct numerical simulation. All fields are expanded in Fourier series, the switch between the physical and Fourier space is performed by fast Fourier transformations \cite{schumacher2007asymptotic} using the software package P3DFFT \cite{pekurovsky2012p3dfft}. The simulation domain is decomposed into pencils and simulation code is parallelized with the Message Passing Interface. Time advancement is done by a second-order predictor-corrector scheme. The same time integration technique\textcolor{red}{\sout{s}} is used for tracking the Lagrangian tracer particles and calculating the Lyapunov exponents \cite{gotzfried2019comparison}. Interpolation to switch between Lagrangian and Eulerian descriptions are done tri-linearly.  
\begin{figure*}
    \centering
    \includegraphics[width=\linewidth]{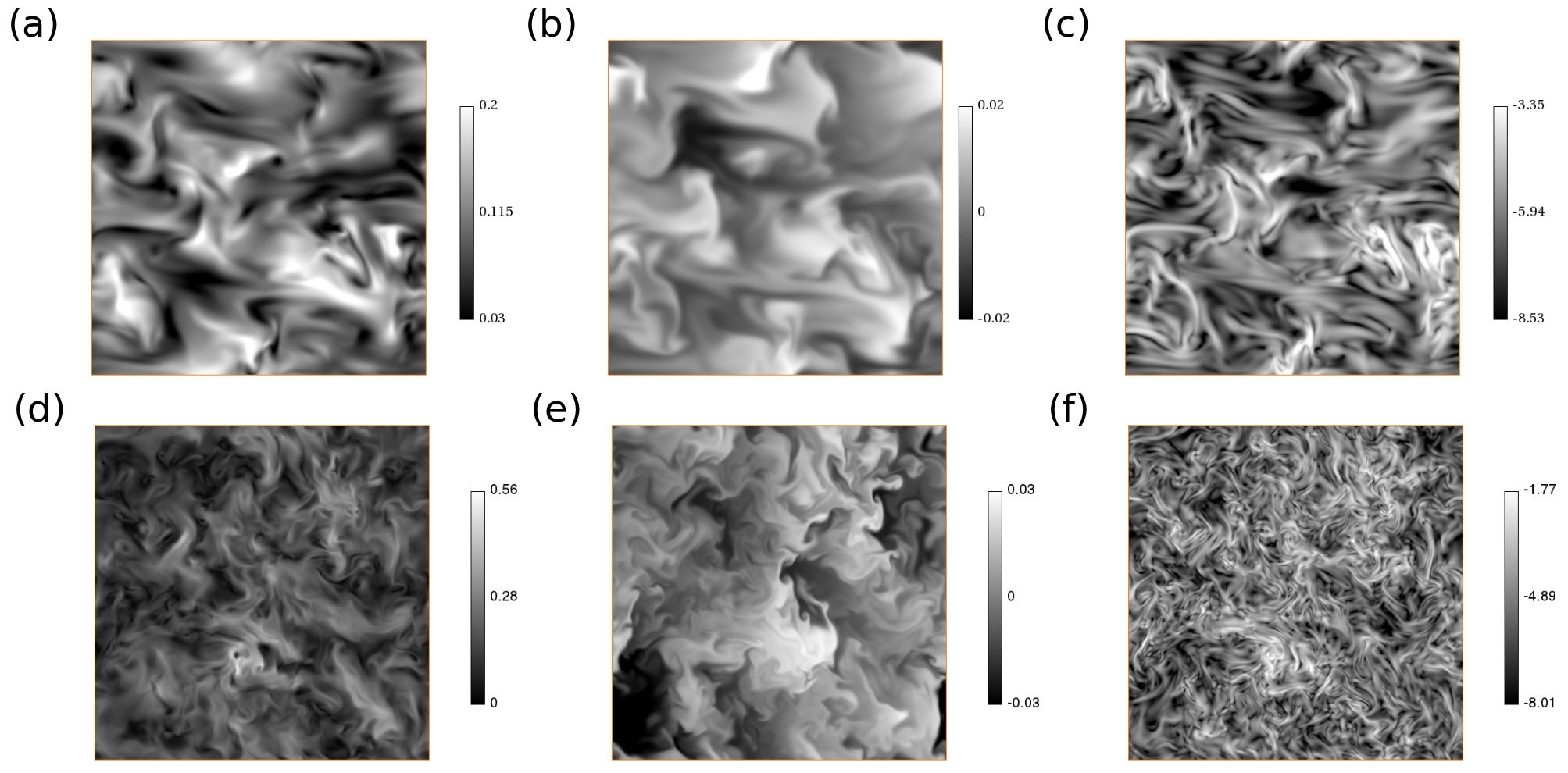}
    \caption{Contour plots of two-dimensional slice cuts of (a) the magnitude of the velocity field, (b) the supersaturation field, and (c) the logarithm of the dissipation rate of turbulent kinetic energy. It is given by $\epsilon = \nu S_{ij}S_{ij}$,  $S_{ij}=(\partial_j u_i + \partial_i u_j)/2$ with the rate-of-strain tensor $S_{ij}$. Data are for the simulation run 2 with $L=256$ mm at $t = 0.045 \tau_L$. The lower panels (d)--(f) shows corresponding plots for simulation run 3 with $L=512$ mm at $t = 0.032 \tau_L$. See also Tab. \ref{table:parameter_study}. The characteristic  unit time unit, the large-scale eddy turnover time $\tau_L$ is specified in Sec. II B, see also Tab. I.}
    \label{fig:eps_u_magn}
\end{figure*}
\begin{figure*}
    \centering
    \includegraphics[width=\textwidth]{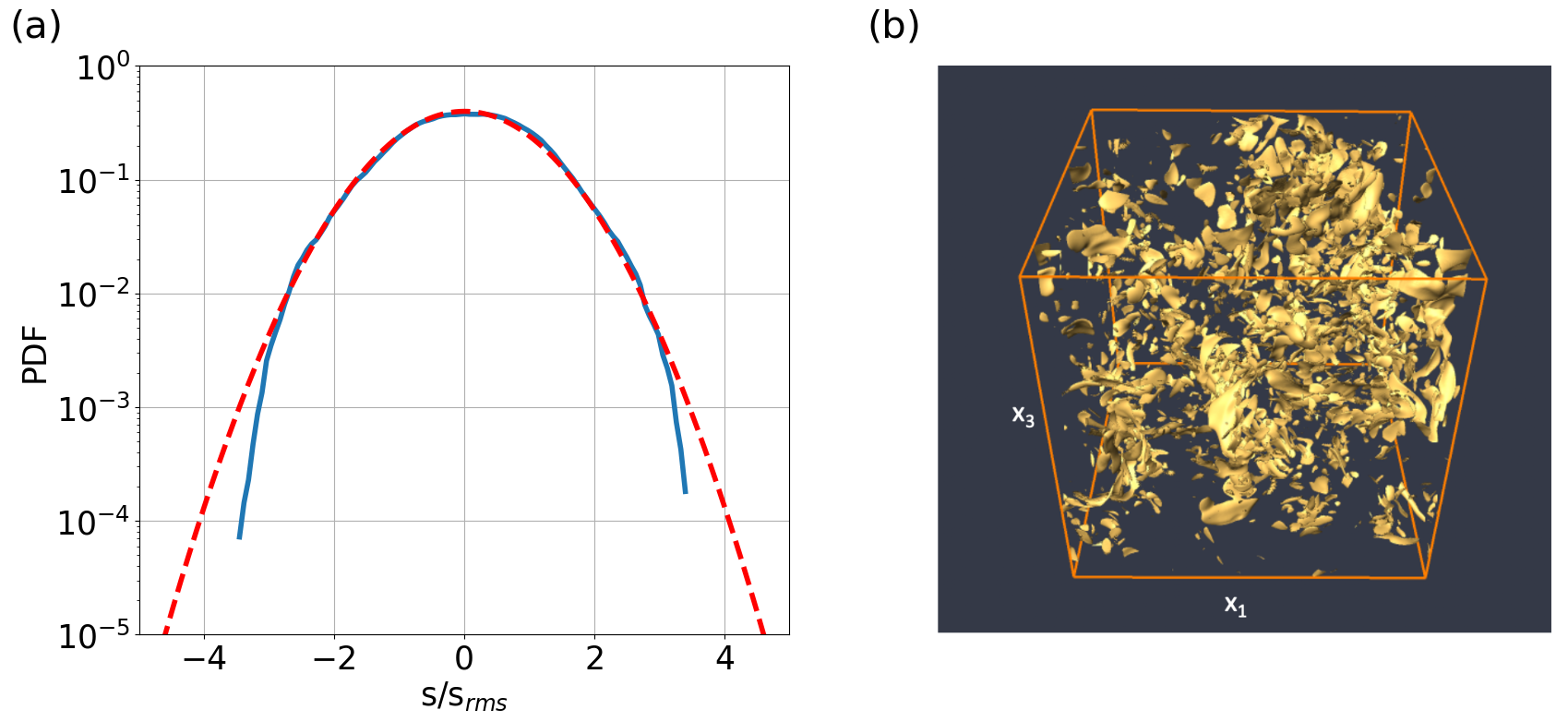}
    \caption{(a) Sub-Gaussian PDF of rescaled supersaturation field taken from Eulerian approach (blue line) and Gaussian PDF for reference. (b) Instantaneous snapshot of isosurfaces of the scalar dissipation rate field $\epsilon_s({\bm x},t)$ which is defined in eq. \eqref{eq:diss_def}. The level of the isosurfaces is $\log_{10}\epsilon_s=-3.3$ in this plot.}
    \label{fig:sub_gaus_and_iso_diss}
\end{figure*}

We vary the Reynolds number and thus the degree of turbulent mixing by an increase of the box size $L$ in the volume forcing term in the Navier-Stokes eqns. \eqref{eq:ns_final}. This forcing is defined such that at each time step a fixed amount of turbulent kinetic energy at a rate $\epsilon_{\rm in}$ is injected \cite{schumacher2007asymptotic}. This term is implemented in the Fourier space,
\begin{equation}
    \label{eq:forcing}
    \mathscr{F}\{{\bm f}({\bm x},t)\} = \epsilon_{\rm in} \frac{\hat{{\bm u}}({\bm k},t)}{\sum_{{\bm k}_f\in K} |\hat{\bm u}({\bm k}_f,t)|^2} \delta_{{\bm k},{\bm k}_f}\,.
\end{equation}
The subset of driven Fourier modes is given $K = \{{\bm k}_f=(2\pi/L)(\pm 1,\pm 1, \pm 2)\}$ plus permutations of wavevector components. Here, $\epsilon_{\rm in}$ is the energy injection rate that prescribes the dissipation of turbulent kinetic energy, i.e.,  $\epsilon_{\rm in} \approx \epsilon$ for the statistically stationary regime. We performed 3 different DNS to investigate how variations of $L$ affect the flow properties. An additional parameter, which quantifies the strength of turbulence, the Taylor microscale Reynolds number is listed in Tab. \ref{table:parameter_study}. It is given by \cite{schumacher2007asymptotic}
\begin{equation}
    R_\lambda = \sqrt{\frac{5}{3\nu \langle \epsilon \rangle}_{V,t}} U^2.
\end{equation}
Table \ref{table:parameter_study} summarizes also other results. With increasing length scale $L$ both Reynolds numbers and the velocity fluctuation magnitude increase, as expected and shown in \cite{kumar2018scale}. Consequently, the large-scale eddy turnover time $\tau_L$ grows as well. The values of $s_{\rm rms}=\langle s^2\rangle_{V,t}^{1/2}$ remains very small, but increases to about $\sim 1\%$ for the biggest domain. We recall that in atmospheric clouds heterogeneous nucleation proceeds such that maximum values of $s$ typically do not exceed a few per cent \cite{yau1996short}. 

Figure \ref{fig:eps_u_magn} visualizes the turbulent dynamics by means of a slice-cut snaphot for runs 2 and 3 of the Tab. \ref{table:parameter_study}. On display are the velocity magnitude (or $E_k$) to the left, the passively mixed supersaturation field, and the logarithm of the kinetic energy dissipation field. It is clearly demonstrated how the complexity of the flow increases when the Reynolds number grows. The kinetic energy dissipation rate field develops finer striated high-amplitude shear layers. They are connected to stretching and compression regions in the Lagrangian picture which will be further investigated in Sec. V by means of the FTLEs.

Figure \ref{fig:sub_gaus_and_iso_diss}a demonstrates probability density function (PDF) for supersaturation field for run 2 which has Gaussian shape in the core and sub-Gaussian tails. This statistics agrees with passive scalar DNS by \citet{celani2001fronts}. Figure \ref{fig:sub_gaus_and_iso_diss}b displays isosurfaces of the instantaneous scalar dissipation rate field of the supersaturation. This quantity is given by
\begin{equation}
\epsilon_s({\bm x},t)= \frac{1}{\rm Re\, Sc} ({\bm \nabla} s({\bm x},t))^2\,.
\label{eq:diss_def}
\end{equation}
We see that the isosurfaces of the scalar dissipation rate at the chosen isolevel are mostly of smaller size, plate-like and not stretched out and curved as it would be the case for turbulent mixing at very large Schmidt numbers, see e.g. \citet{Kushnir2006}. The reason is that the chaotic stirring of the scalar at sub-Kolmogorov scales is absent for the present Schmidt number Sc $\sim 1$. In other words, no viscous-convective range exists for the passive scalar in the present case which is  established between the Kolmogorov length $\eta_K$ and the diffusive length of the passive scalar, the Batchelor length $\eta_B=\eta_K/\sqrt{\rm Sc}$. This will have implications for the prospective application of the aggregation model of passive scalars \cite{Villermaux2003,duplat2008mixing} which we will discuss further below in the text.

\section{Cloud water droplet dynamics and statistics}
\begin{figure*}
    \centering
    \includegraphics[width=\linewidth]{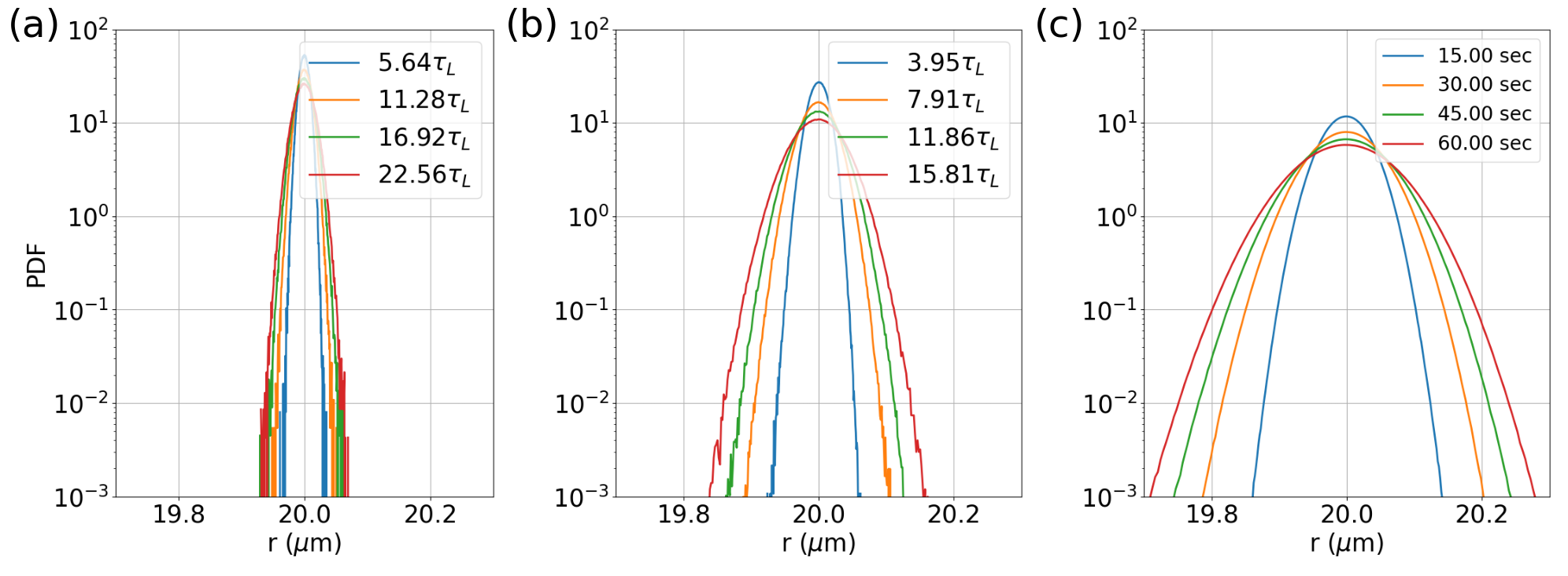}
    \caption{Droplet size distribution at different time instants for computational domains with (a) $L = 128$ mm (run 1), (b) $L = 256$ mm (run 2) and (c) $L = 512$ mm (run 3). All times are expressed in units of the corresponding large-scale eddy turnover time $\tau_L$ which are provided in Tab. \ref{table:parameter_study}.}
     \label{fig:size_distr1}
\end{figure*}

\subsection{Cloud water droplets as Lagrangian tracers}
Each individual cloud water droplet was advected by statistically steady turbulent flow. It is seen from eq. (\ref{eq:radius}) that the supersaturation field directly impacts on the size distribution of cloud droplets. Figure \ref{fig:size_distr1} displays the droplet size distributions for runs 1, 2 and 3 taken at different time instants. The probability density function broadens in all cases with progressing time, but remains in a Gaussian shape without developing extended tails in the cloud bulk.

As seen in Tab. \ref{table:parameter_study}, a larger domain size results in a larger velocity fluctuation magnitude. Consequently, the fluctuations of the supersaturation field are enhanced, see again Tab. \ref{table:parameter_study}. As a consequence and as expected, the cloud droplet size distribution broadens much faster in time. Note that we provide the instants in terms of the corresponding large-scale eddy turnover time scale $\tau_L$, which itself becomes larger as the domain size increases.
\begin{figure*}[!t]
    \centering
    \includegraphics[width=\linewidth]{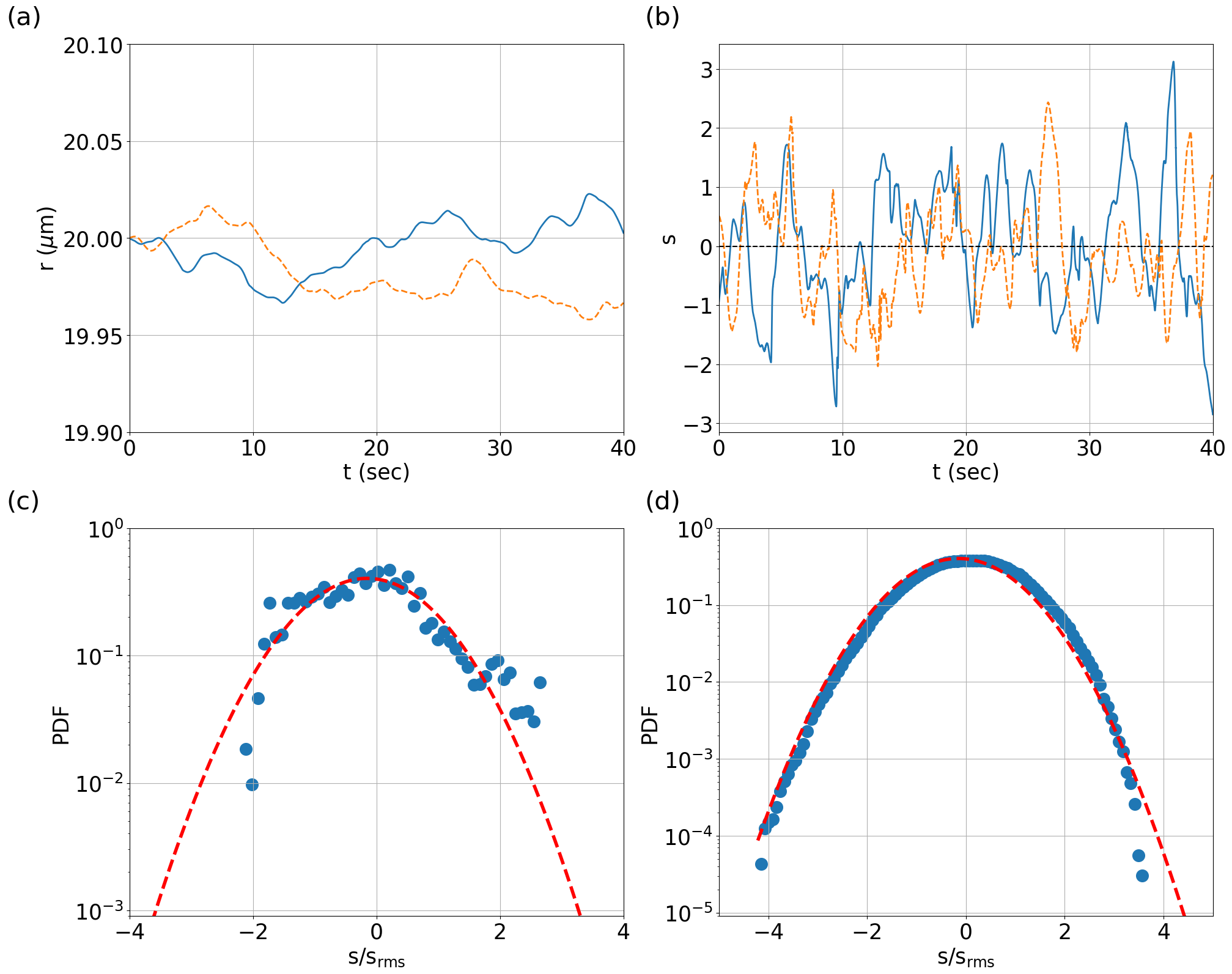}
    \caption{Time evolution of two individual droplet radii in (a) and the corresponding supersaturation field in (b). Solid and dashed lines correspond to a particular particle. Time is given in seconds. (c) Probability density function (PDF) of the rescaled supersaturation field, $s/s_{\rm rms}$ experienced by the chosen individual particle for the whole calculation period and (d) PDF of rescaled supersaturation field for all particles averaged in time. The dashed line shows the corresponding Gaussian PDF for comparison.}
    \label{fig:individual_tracers}
\end{figure*}

A more detailed analysis for individual cloud droplets was performed subsequently. Therefore, the Lagrangian droplets data of a few randomly selected particles, which are initially far enough separated from each other, were written out at each time integration step. The first plot of Fig. \ref{fig:individual_tracers}a demonstrates the variation of the droplet radius for two selected particles for a physical time of 2 minutes. The second plot of the figure in panel (b) shows the corresponding value of the supersaturation field $s({\bm X}_p,t)$ at the particle position ${\bm X}_p(t)$. It can be clearly seen that time intervals with positive supersaturation ($s>0$) correspond to the growth periods of cloud droplets. Droplets shrink in sub-saturated air ($s<0$) \cite{celani2005droplet}.

According to statistically steady homogeneous isotropic bulk turbulence, we expect that PDF of $s({\bm X},t)$ converges to the Gaussian distribution in the core with sub-Gaussian tails. Particles, which are advected in the turbulent volume are affected by differently long intervals of super- and subsaturation. Figure \ref{fig:individual_tracers}c confirms this point; we reproduce the Gaussian statistics in the core with the sub-Gaussian tails in the Lagrangian frame which agrees consistently with the Eulerian analysis which was dispayed in Fig. \ref{fig:sub_gaus_and_iso_diss}.

The evaporation and condensation processes of the cloud droplets depend on the character of corresponding mixing regime as shown in several field measurements and controlled laboratory experiments, see e.g. refs. \cite{Beals2015,Chandrakar2016,Chandrakar2020}.  As discussed already above in the introduction, the mixing regime is defined by a Damk\"ohler number. For cloud droplet evaporation, we take $\textrm{Da}_d$ in (\ref{eq:param}). In the homogeneous mixing regime for $\textrm{Da}_d\ll1$, droplets evaporate with almost equal rate while in the inhomogeneous mixing regime evaporation rates of different droplets are different, i.e., some droplets can evaporate completely while others are not evaporated at all. Here, the cloud droplet evaporation is a homogeneous mixing process which explains the Gaussian distribution of the cloud droplet radii. The PDFs are without far-tails as they were seen in Kumar et al. \cite{Kumar2014} for the edge of the cloud.
The second thermodynamic process of interest, the saturation relaxation of the water vapour content, is parameterised by the second Damköhler number $\textrm{Da}_s$ in eq. (\ref{eq:param}). The corresponding relaxation time $\tau_s$ of the passive supersaturation field is of the order of $\tau_L$. All three runs fall into the category of  inhomogeneous mixing with $\textrm{Da}_s\gtrsim 1$, which implies that regions of a cloud are saturated slightly faster than they are mixed by the turbulence. The vapour field is stirred by the fluid turbulence, fluctuations of $s({\bm x},t)$ are effectively sustained in a statistically steady state, and the relaxation time scale $\tau_s \lesssim \tau_L$.

\begin{figure}[!hb]
    \centering
    \includegraphics[trim={0cm 0.2cm 0cm 0.3cm},clip,width=0.5\linewidth]{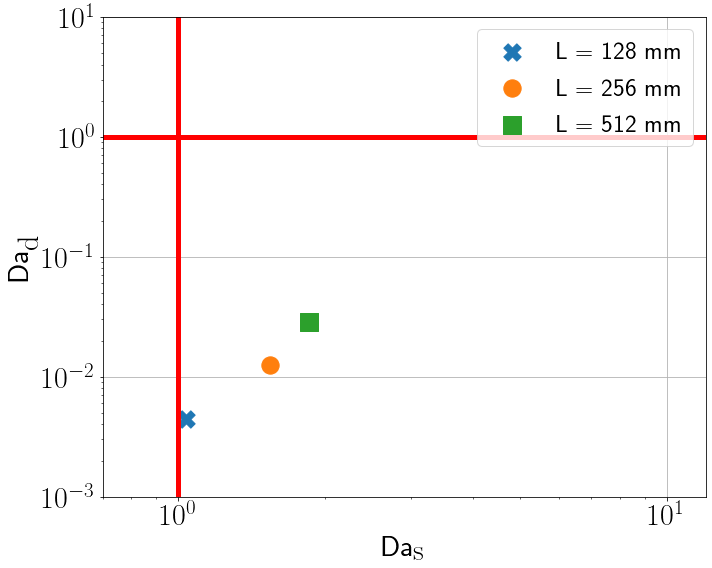}
    \caption{Parameter space which is spanned by the two Damk\"ohler numbers ${\rm Da}_d$ and ${\rm Da}_s$. The operating points of the 3 simulation runs are inserted, see the legend and Tab. \ref{table:parameter_study}. Solid lines show $\text{Da}=1$ for both. They mark the corresponding transition between homogeneous and inhomogeneous mixing. Damk\"ohler numbers greater and smaller than one characterise homogeneous and inhomogeneous mixing regimes, respectively.}
    \label{fig:Da_space}
\end{figure}

Both Damköhler numbers define an operating point in the parameter plane of our stationary model, $({\rm Da}_s,{\rm Da}_d)$. They are shown for the 3 simulations at different lengths $L$ = 128 mm, 256 mm and 512 mm in Fig. \ref{fig:Da_space}. Markers characterise the position of our simulation cases in the Damköhler number space. The solid vertical and horizontal lines characterise the transition from homogeneous to inhomogeneous mixing for both ``reaction" processes at $\textrm{Da}=1$. 

\subsection{Effect of gravity settling and magnitude of constant $A_1$}
\begin{table*}
    \begin{ruledtabular}
        \begin{tabular}{lccc}
            Run &  $A_1$ & $r_0$ & Inertia and gravity\\
            & $m^{-1}$& $\mu$m &\\ \hline
            2a & $5\cdot 10^{-4}$ & 10 & - \\
            2b & $5\cdot 10^{-4}$ & 10 & + \\
            2c & 0.2 & 20 & + \\
            2d & 0.2 & 20 & - \\
            2e & $5\cdot 10^{-4}$ & 20 & - \\
            2f & $5\cdot 10^{-4}$ & 20 & + \\
        \end{tabular}
    \end{ruledtabular}
    \caption{Parameters of the additional simulation runs 2a to 2f which have the same setting for the turbulent velocity as run 2. These runs have either a smaller constant $A_1$ in the transport equation for the supersaturation field or inertia effects included or both. Plus mean included; minus means neglected.}
    \label{table:additional_runs}
\end{table*}
In the previous subsection, gravitational settling and droplet inertia were neglected. Their impact is studied now. Similar to previous works \cite{Kumar2013,Lanotte2009,sardina2015continuous,Gotoh2016}, we generalize our droplet dynamics model in the following to one for inertial particles. The Lagrangian equations of motion are then extended to
\begin{subequations}
    \begin{align}
        \frac{d {\bm X}_i}{dt} &= {\bm V}_i\,,\\
        \frac{d {\bm V}_i}{dt} &= \frac{1}{\tau^{(i)}_{p}}\left[{\bm u}({\bm X}_i,t) -{\bm V}_i\right] - g {\bm e}_z,
    \end{align}
    \label{eq:settled_part}
\end{subequations}
for $i = 1,\dots,N_{0}$. Here, ${\bm X_i}$ and ${\bm V_i}$ are position and velocity for $i$-th particle respectively. Time $\tau^{(i)}_{p}=2 \rho_L r_i^2/(9\rho \nu)$ is the  particle response time of the $i$-th droplet and $g$ the acceleration due to gravity. To this end, we performed six additional DNS in the same setting as run 2 of Tab. \ref{table:parameter_study}, but with new equations for the particle dynamics, see eqns. \eqref{eq:settled_part}. Different combinations of the magnitude of $A_1$, different initial radii, and droplet inertia are covered by these runs. They are  summarized in Tab. \ref{table:additional_runs}.

The particle response time $\tau_p$ depends on its radius; for particles with a radius $r \approx 20$ $\mu m$ the response time is $\tau_p \approx 0.0045$ sec. Together with the large-scale eddy turnover time $\tau_L = 2.64$ sec as the characteristic velocity field time-scale (for runs 2a to 2f), the Stokes number ${\rm St}=\tau_p/\tau_L\ll 1$. This implies that the Stokes friction term, i.e., the first term on the right hand side of the particle velocity eq. (\ref{eq:settled_part}b), will remain subdominant in comparison to the gravitational settling term, the second term on the right hand side. Furthermore, a second initial droplet radius of $r_0=10$ $ \mu m$ was taken which leads to different growth rates. Here, inertia effects are even smaller; the radius is however still large enough to neglect curvature and hygroscopicity effects for the growth by vapor diffusion of the water droplets \cite{yau1996short}.

\begin{figure}
    \centering    \includegraphics[width=\linewidth]{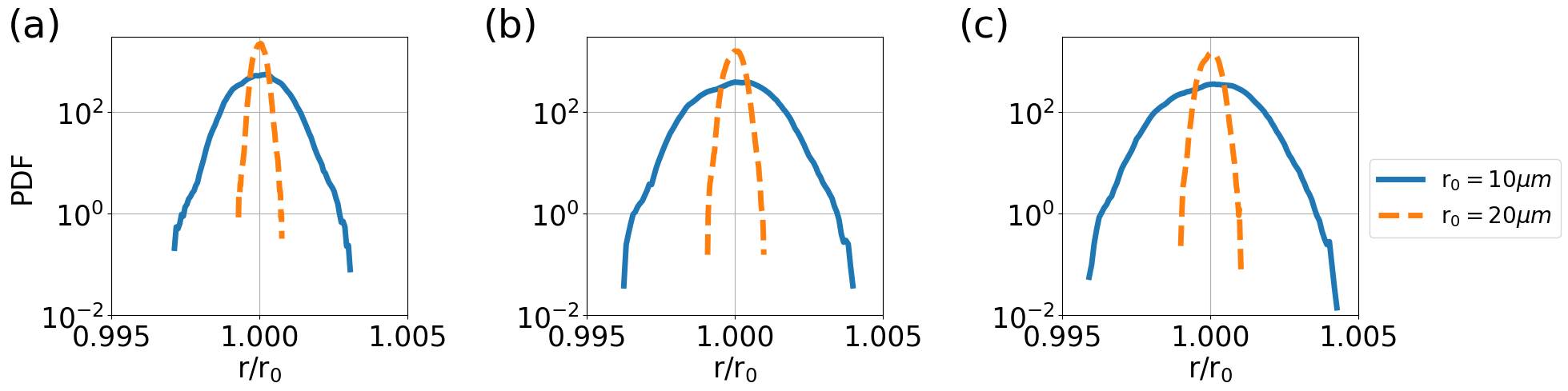}
    \caption{The impact of different initial sizes of the cloud droplets on the droplet size distribution. Panel (a) is taken at $t = 1$ sec, (b) at $t = 2$ sec, and (c) $t = 3$ sec. Data correspond to runs 2a and 2e in Tab. \ref{table:additional_runs}.}
    \label{fig:vary_radius}
\end{figure}

\begin{figure}[b]
\includegraphics[width=0.5\linewidth]{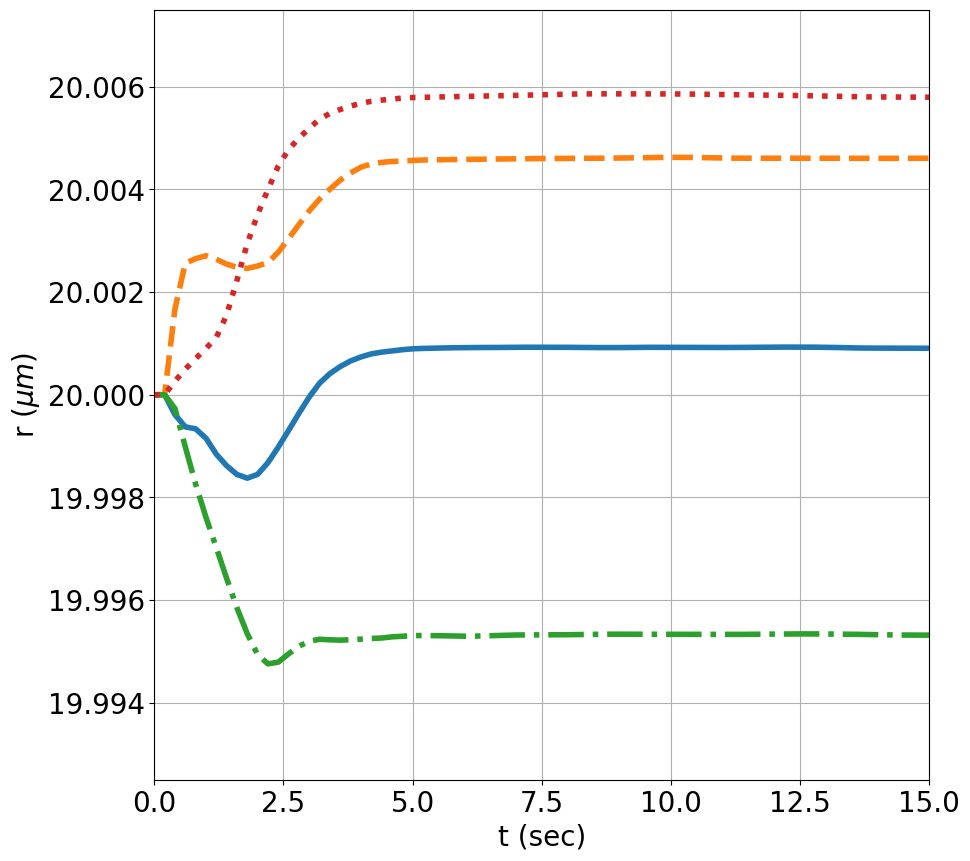}
\caption{Droplet radius versus time for four randomly chosen individual droplets in run 2e from Tab. \ref{table:additional_runs}.}
\label{fig:evap_real_a1}
\end{figure}

As seen in Fig. \ref{fig:vary_radius}, the smaller initial droplet size leads to a faster {\em relative} growth and shrinking of droplets and thus to a faster relative broadening of the droplet size distribution. Both runs (2a and 2e) were conducted at the corresponding realistic  value of $A_1=5\cdot 10^{-4}$ \cite{Celani2008,sardina2015continuous}. Thus even at this small $A_1$, a small change of the droplet size is observed for the short mixing process. After approximately 2 large-scale eddy turnover times, which corresponds to $t\approx 5.3$ secs, the mixing process is ceased and the droplet size distribution reaches a steady state, as shown in Fig. \ref{fig:evap_real_a1} where we plot the radius of 4 selected droplets as a function of time.
\begin{figure}
    \centering
    \includegraphics[width=\linewidth]{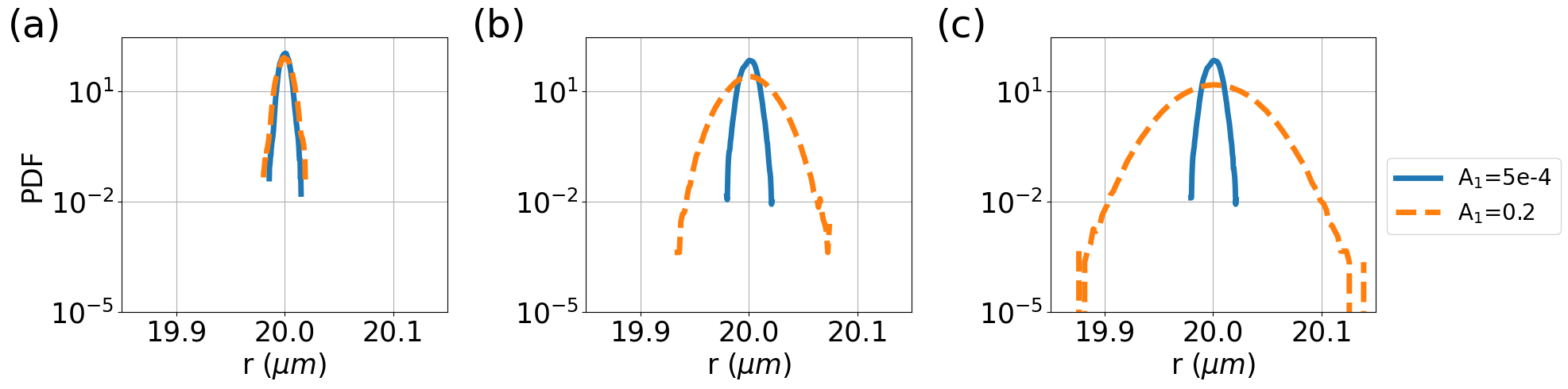}
        \caption{Comparison of the time evolution of the cloud droplet size distribution for non-enhanced prefactor, $A_1=5\cdot10^{-4}$, (solid line) and enhanced factor, $A_1=0.2$, (dashed line)  for cloud droplets  with initial radius $r_0=20 $ $\mu m$. These are runs 2d and 2e from Tab. \ref{table:additional_runs}. Panel (a) is taken at $t = 1$ sec, (b) at $t = 5$ sec, and (c) $t = 15$ sec.}
    \label{fig:size_distr_with_and_without_g_20}
    \label{fig:diff_a1}
\end{figure}
\begin{figure}[b]
    \centering
    \includegraphics[width=\linewidth]{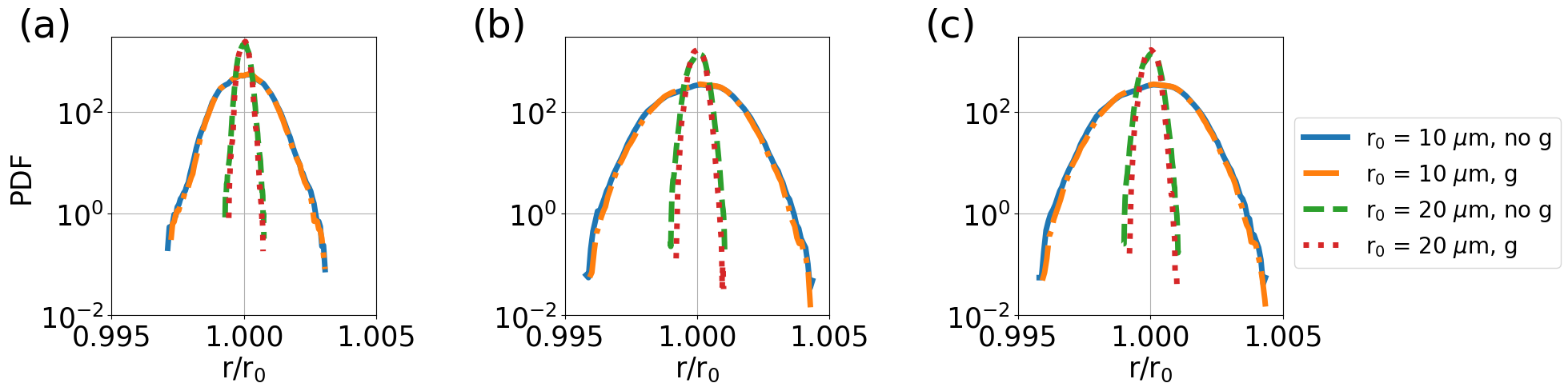}
        \caption{Comparison of the time evolution of the cloud droplet size distribution for non-enhanced prefactor, $A_1=5\cdot10^{-4}$, for cloud droplets with dynamics as gravity settling particles (dash-dotted and dotted lines) and as Lagrangian tracers (solid and dashed lines) for droplets with initial radius $r_0=20$ $\mu m$ (dash-dotted and dashed lines) and $r_0=10$ $\mu m$ (solid and dashed lines). These are runs 2a, 2b, 2e and 2f from Tab. \ref{table:additional_runs}. Panel (a) is taken at $t = 1$ sec, (b) at $t = 5$ sec, and (c) $t = 15$ sec.}
\end{figure}

The first of the two forcing terms, $A_1 u_z$, in eq. \eqref{eq:s_full} keeps the supersaturation field in a statistically steady state; the second term was found to have a very small impact on the dynamics only. With increasing box size and Reynolds number both terms would gain magnitude and thus enhance the level of fluctuations of the supersaturation field. The present box sizes are too small, such that we decided to enhance the prefactor artificially to keep the supersaturation fluctuations at about $s_{rms}\approx 2\%$. Increasing $A_1$ results in an expected faster broadening PDF of droplet radii as can be seen from Fig. \ref{fig:diff_a1}. Here, we compared identical simulations with an enhanced $A_1=0.2$ to ones with the realistic value for the bulk of a warm cloud, $A_1\approx 5\times 10^{-4}$.

Figure \ref{fig:size_distr_with_and_without_g_20} reports a further effect, the longer-term impact of droplet inertia on the broadening of the droplet size distribution for two different initial radii and different $A_1$. For the initial radius of $r_0 = 10$ $ \mu m$, the additional impact of inertia, i.e., gravitational settling, remains very small such that the droplet size distributions for droplets with and without inertia collapse almost perfectly for all times reported here and both $A_1$. This holds even for the tails. In case of initial radius of $r_0 = 20$ $ \mu m$, there is a small effect of the gravitational settling term  observable. The gravitational settling term causes a partial decoupling of the droplet tracks from the velocity field and thus from the simultaneously stirred filaments of the supersaturation field. The vertical motion of these bigger droplets is slightly stronger in comparison to the horizontal one. This leads to slightly sparser tails of the droplet size distribution which is visible in all three panels for all three times. The effect becomes visible for the larger $A_1=0.2$; it is not detectable for the realistic value of $A_1$.

\section{Finite-time Lyapunov exponents and supersaturation}

In the Lagrangian approach, the turbulence can be studied along the lines of dynamical systems theory. The deformation of the fluid element assigned with each particle in the turbulent flow can be used to obtain and monitor the local stretching and compression in the flow along droplet trajectories. This information is obtained by means of the finite-time Lyapunov exponents (FTLE). They are denoted as $\lambda_i$ with $i=1,2,3$. Applying the gradient of eq. \eqref{eq:lagr_tracer} with respect to initial condition $X_{i0}$, gives \cite{johnson2015large,pikovsky2016lyapunov}
\begin{equation}
\label{eq:grad_of_tracer}
\frac{d M_{ij}(t)}{dt} = J_{ik}(t)M_{kj}(t) \quad\mbox{with}\quad
M_{ij}(0) = \delta_{ij},        
\end{equation}
where $M_{ij}=\partial X_i/\partial X_{j0}$ is the deformation tensor, $J_{ij}=\partial u_i/\partial X_{j0}$ the Jacobian of the velocity field, and $\delta_{ij}$ the Kronecker delta. Integrating eq. \eqref{eq:grad_of_tracer} with respect to time $t$, which is taken as a multiple integer of the step width $\Delta t$, one obtains for the $l$-th particle at the position $X_l$ at time $t=n\Delta t$ a tensor $M_{ij}$ that is given by
\begin{figure*}
    \centering
    \includegraphics[width=\linewidth]{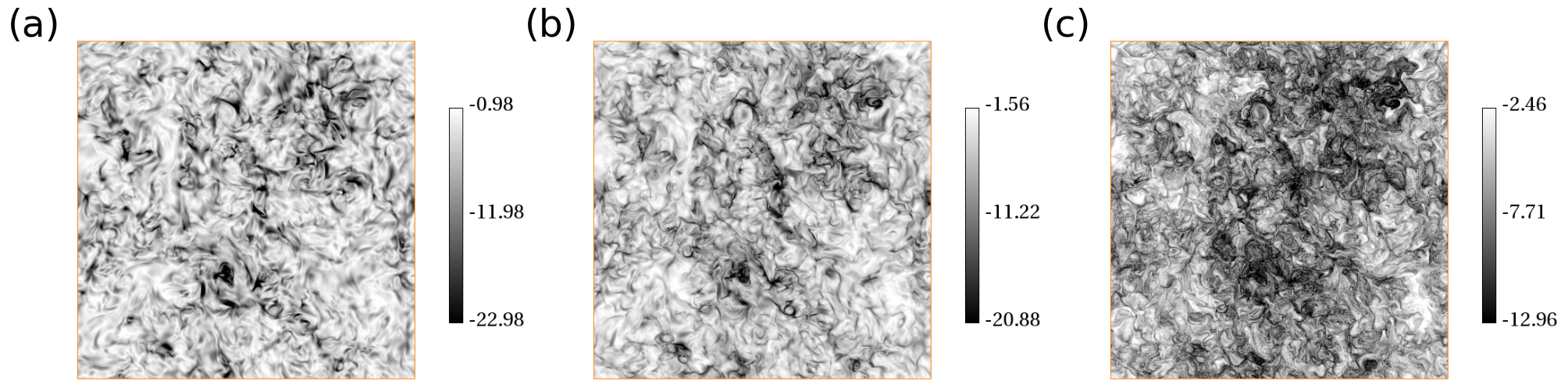}
    \caption{Contour plots of horizontal slice cuts of $\lambda_3({\bm x},t)$ at different times in units of the large-scale eddy turnover time. (a) t = 0.03$\tau_L$, (b) t = 0.09$\tau_L$, and (c) t = 0.32$\tau_L$. The magnitudes are given by the corresponding color bars. Data are for run 3.}
    \label{fig:lambda_3}
\end{figure*}
\begin{equation}
    M_{ij}^n = \left[ \delta_{ik} + J_{ik}\big|_{{\bm X}_{i}^n} \Delta t\right]M_{ij}^{n-1}.
\end{equation}
\begin{figure}[!b]
    \centering
    \includegraphics[width=\linewidth]{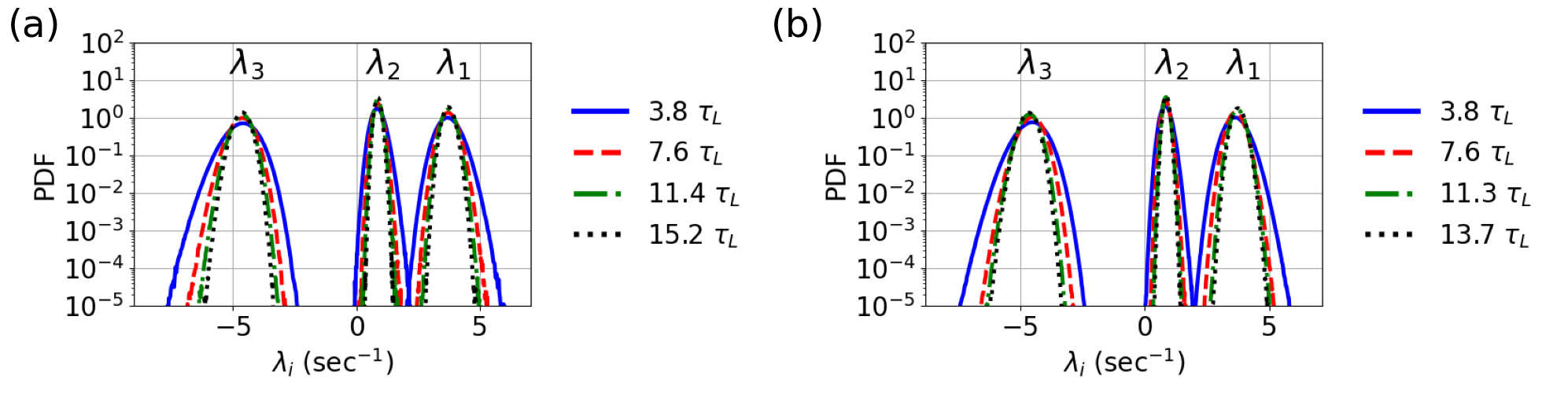}
    \caption{Probability density functions of the finite-time Lyapunov exponents at different times in units of the large-scale eddy turnover time $\tau_L$.  (a) Run 2 with $L =$ 256 mm and (b) run 3 with $L =$ 512 mm.}
    \label{fig:pdf_FTLE}
\end{figure}

For each time step $n$, a QR-decomposition of the deformation tensor $M_{ij}$ is performed, i.e.,
\begin{equation}
    \hat{M}^n = \hat{Q}^n \hat{R}^n\,,
\end{equation}
where $\hat{Q}^n$ is an orthogonal matrix, $\hat{R}^n$ the upper-triangle matrix. Since $\hat{Q}^n$ involves only rotations and reflections, the matrix $R^n$ contains information about stretching and compression of the corresponding fluid element, which is assigned with a Lagrangian particle. FTLEs are obtained from the time-averaged exponential growth or decay of the diagonal elements, $R^n_{ii}$ (note that no Einstein summation rule is applied here). In detail, the exponents are given by
\begin{equation}
    \lambda_i = \frac{1}{n \Delta t}\sum_{j=1}^{n}\ln{\left|R_{ii}^j \right|}\,.
\end{equation}
\begin{figure}[!b]
    \centering
    \includegraphics[width=0.45\linewidth]{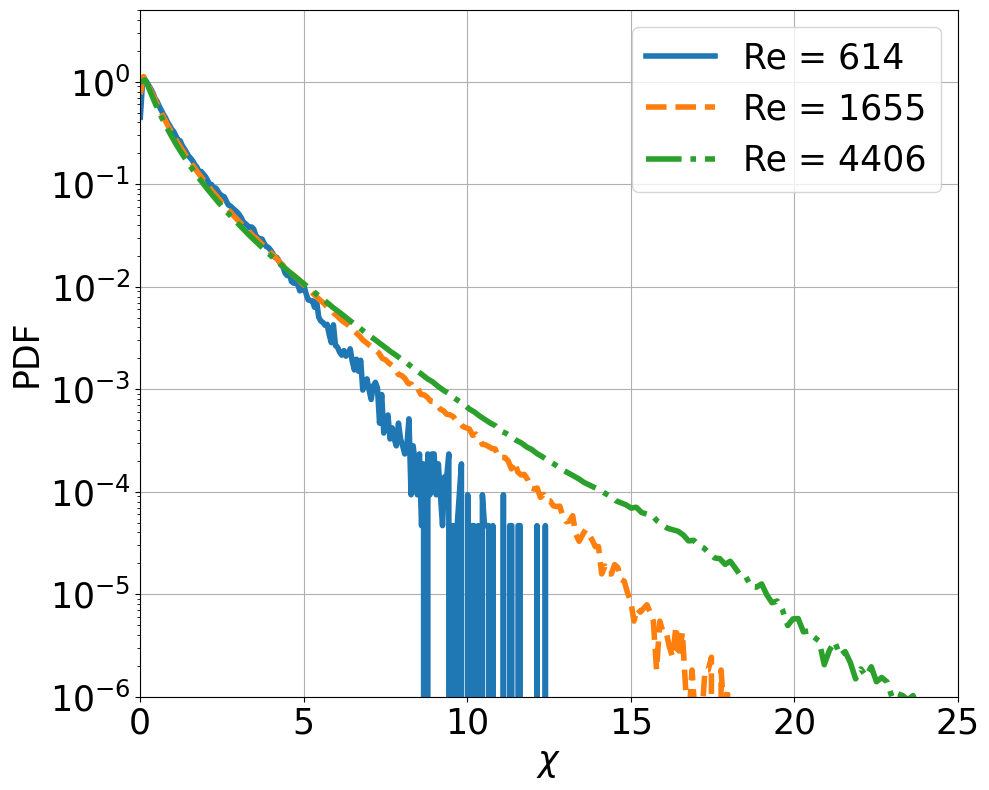}
    \caption{Probability density function of the magnitude of the gradient of the supersaturation field, $\chi=|{\bm \nabla}s|/|{\bm \nabla}s|_{\rm rms}$ for runs 1 (solid line), 2 (dashed line), and 3 (dash-dotted line). We indicate the corresponding Reynolds numbers in the legend.}
    \label{fig:pdf_s_grad_s}
\end{figure}
\begin{figure*}[!ht]
    \centering
    \includegraphics[width=\linewidth]{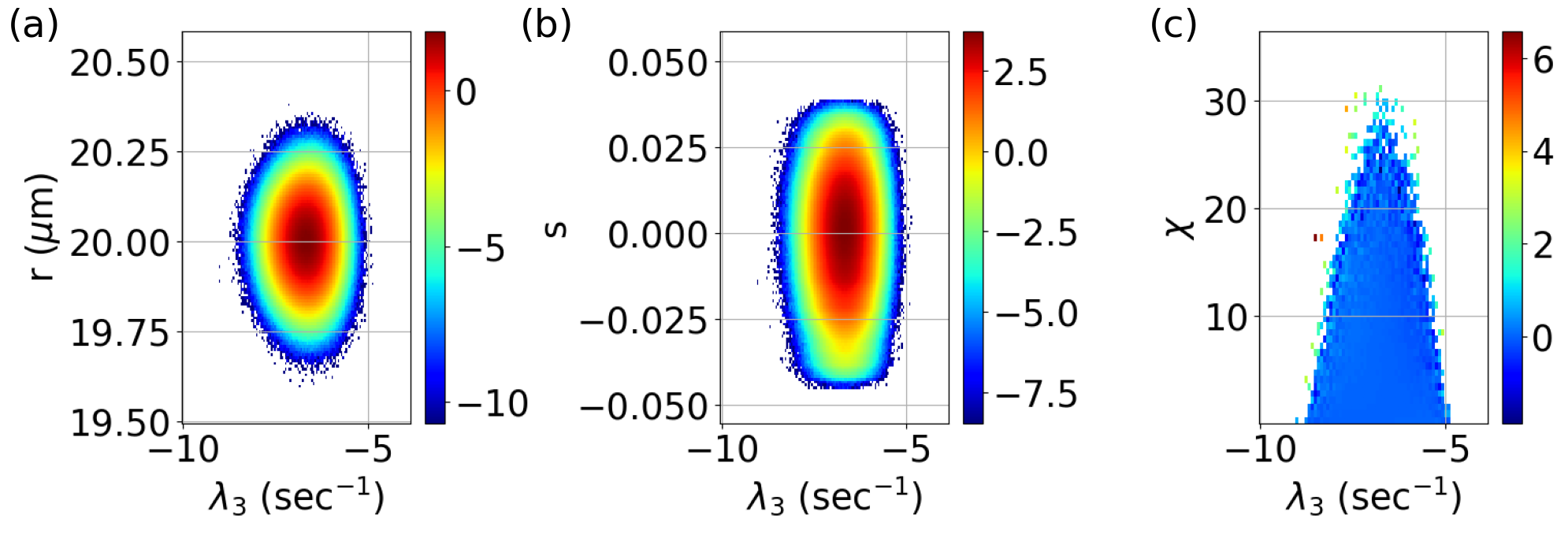}
    \caption{Joint probability density functions of the third FTLE with (a) the droplet radius, $p(\lambda_3,r)$, (b) the supersaturation field at the particle positions, $(p(\lambda_3,s)$, and (c) the normalized gradient magnitude of the supersaturation field, $\tilde{p}(\lambda_3,\chi)$, see \eqref{norma} for specific definition. Data are for run 3 at $t=19.4 \tau_L$, see also Fig. \ref{fig:pdf_FTLE}(b).}
    \label{fig:joint_pdf}
\end{figure*}
The FTLEs are then ordered, $\lambda_1 > \lambda_2 > \lambda_3$ for each reference tracer. Since we consider an incompressible flow, the sum of FTLEs follows to
\begin{equation}
    \lambda_1 + \lambda_2 + \lambda_3 = 0.
\end{equation}
\begin{table*}
    \begin{ruledtabular}
        \begin{tabular}{lccc}
            Run &  $\langle\lambda_1\rangle\: :\: \langle\lambda_2\rangle\: : \:\langle\lambda_3\rangle$\\ \hline
            1 & $3.8 \: : \:0.9\: : \:-4.7$ \\
            2 & $3.7 \: : \:0.9\: : \:-4.6$ \\
            3 & $3.7 \: : \:0.8\: : \:-4.5$ \\
        \end{tabular}
    \end{ruledtabular}        
    \caption{Ratio between mean value of FTLE for runs from Table \ref{table:parameter_study}. The average is taken over all Lagragian droplet tracks.}
    \label{table:ftle-ratio}
\end{table*}
The first FTLE $\lambda_1$ is always positive and characterizes local stretching, the intermediate $\lambda_2$ can be either negative or positive. The third exponent, $\lambda_3$, is always negative and describes the local compression of the fluid element. This exponent characterizes the pile-up of scalar concentration by stirring which can lead to an aggregation of filaments. The evolution of the compression field $\lambda_3({\bm x},t)$ is displayed in Fig. \ref{fig:lambda_3}. With progressing time the contours gets increasingly finer in scale and convoluted.  Figure \ref{fig:pdf_FTLE} shows the PDFs of all three FTLEs for different moments of time. As can be seen, the PDFs of the FTLEs converge to Gaussian distributions for later times, $t\gtrsim 5\tau_L$. The ratio between mean values of the three FTLEs is seen in Tab. \ref{table:ftle-ratio}. With increasing Reynolds number the magnitude of all three mean values decreases slightly. The ratio is of approximately the same size as in \citet{johnson2015large}. The rates of convergence to the Gaussian case in units of $\tau_L$ are approximately the same for the investigated runs, as seen in the figure. We have verified this relaxation by means of the skewness and flatness factors of the PDF($\lambda_3$) which converge to 0 and 3, respectively. The range of Taylor microscale Reynolds numbers $R_{\lambda}$ in this study is limited; it basically ends where the range of the studies of \citet{bec2006lyapunov} starts; it is much smaller than in \citet{johnson2015large}. This might be the reason of why the PDFs for all $\lambda_i$ become symmetric in our study and do not show the asymmetric form which would follow from the large deviation theory (once the turbulent flows are highly intermittent in the inertial cascade range).

The regions of high compression rates, as characterised by high absolute values of $\lambda_3$, describe regions of high turbulent mixing (which means stirring plus diffusion). Thus regions with high magnitudes of $\lambda_3$ should also be associated with regions of high spatial variations of the supersaturation field $s({\bm x},t)$, which is probed by the magnitude of the supersaturation gradient or the corresponding scalar dissipation rate, see again Fig. \ref{fig:sub_gaus_and_iso_diss}. We display the PDFs of the gradient of the supersaturation field (normalized by its rms value) in Fig. \ref{fig:pdf_s_grad_s}. The distributions are highly intermittent, as expected. Tails are increasingly extended with growing Reynolds number Re. The far-tail regions are taken as those, where $\chi=|{\bm \nabla}s|/|{\bm \nabla}s|_{\rm rms}\ge 5$. 

Figure \ref{fig:joint_pdf} shows different joint probability density functions (JPDFs). These are $p(\lambda_3, r)$ in panel (a), $p(\lambda_3, s)$ in panel (b), and $p(\lambda_3,\chi)$ in panel (c). Panels (a) and (b) show the typical elliptically shaped contours of the joint PDFs since both distributions are Gaussian. Panel (c) replots the JPDF normalized by the single quantity marginal PDFs
\begin{equation}
    \tilde{p}(\lambda_3, \chi) = \frac{p(\lambda_3,\chi)}{p(\lambda_3)\, p(\chi)}\,.
    \label{norma}
\end{equation}
When the quantity $\tilde{p}>1$ then the correlation of both statistical quantities is larger as if they were statistically independent. We see that this is exactly the case at the outside of the support of the JPDF, particularly where larger $\lambda_3$ are connected with larger $\chi$. It is expected that this effect becomes stronger when the analysis is moved towards the cloud boundary. 

\begin{figure}
    \includegraphics[width=\linewidth]{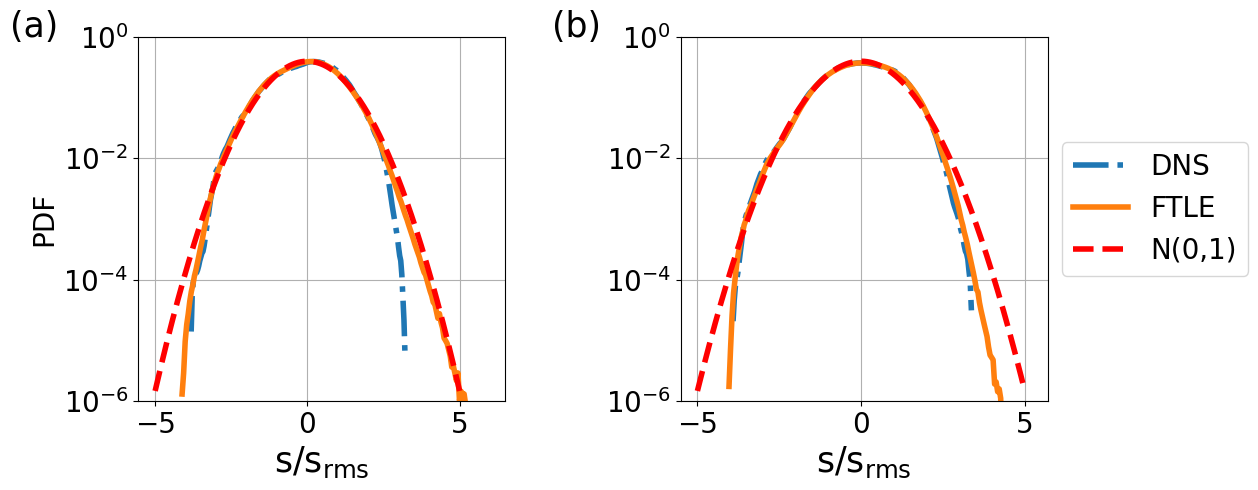}  
    \caption{Probability density functions of supersaturation $s$ for times (a) $t=4.86\tau_L$ and (b) $t=9.75\tau_L$, which are obtained from the FTLEs (solid line) from run 3 via the aggregation model, are compared to the directly evaluated ones (dash-dotted line). The dashed line is a Gaussian PDF.}
    \label{fig:aggr_pdf}
\end{figure}

As mentioned above, the FTLEs are used to monitor stretching and compression regions in the turbulent flow. These regions are connected to changes of the supersaturation field. Thus the PDF of $\lambda_3$ can be connected to the one of $s$. Locally, the growth of a filament of the supersaturation can be approximated  effectively from an one-dimensional advection by compressive strain \cite{gotzfried2019comparison}
\begin{equation}
    S(t)\approx S(t_0) \; \textrm{exp} \left[\lambda_3(t)(t-t_0)\right],
    \label{eq:aggreg_model_0}
\end{equation}
where we define $S({\bm x},t) = s({\bm x},t) + 1$ for convenience. This filament aggregation model \cite{villermaux2003mixing} is applicable in the crossover region starting from the small-scale end of the inertial range into the viscous range below the Kolmogorov length $\eta_K$. This viscous-convective range of passive scalar turbulence is well developed only when $Sc \gg 1$ \cite{gotzfried2019comparison}. Here, the viscous-convective range is narrow (if existing at all). In the crossover range up to scales of a few Kolmogorov lengths $\eta_K$ the velocity field should however still be sufficiently smooth such that the framework is applicable. Time $t_0$ in eq. \eqref{eq:aggreg_model_0} is the initial time of a time lag over which such a local aggregation process proceeds at many places simultaneously in the bulk of a cloud. The original model of \citet{Villermaux2003} is formulated for a conserved passive scalar field that is not subject to additional source terms in contrast to the present situation. We incorporate the relative changes of the compression rates $\lambda_3(t)$ relative to $t_0$ by
\begin{equation}
    S(t) \approx S(t_0) \: \textrm{exp} \left[\frac{\lambda_3(t)}{\lambda_3(t_0)}-1\right].
    \label{eq:aggr_model}
\end{equation}
Figure \ref{fig:aggr_pdf} compares the supersaturation PDF for $s$ obtained from the PDF of the compressive FTLE with scalar distribution obtained in DNS. Therefore, we applied the substitution rule
\begin{equation}
P(\tilde{S})= \int p(\lambda_3,t) \,\delta(\tilde{S}-S(\lambda_3,t)) d\lambda_3\,,
\end{equation}
where $S(\lambda_3,t)$ follows from \eqref{eq:aggr_model}. As seen, the results obtained from FTLE field show a good agreement with the original analysis of the DNS data. Particulalrly for later times, the sub-Gaussian tails of the PDF of the supersaturation field are reproduced. A task for the future work is to check how well this aggregation model will work when we move towards the edge of the cloud where entrainment processes become important and multiple filament foldings might take place.

\section{Summary and outlook}
In the present work, a simple warm cloud mixing model is presented and analysed. Instead of two scalar fields, the water vapor mixing ratio $q_v$ and the temperature $T$, one scalar field, the supersaturation $s$, is used. This supersaturation field contains all required information about water vapor and temperature in the bulk of the cloud and thus determines the evaporation and condensation of cloud droplets \cite{Lanotte2009,sardina2015continuous,fries2021key}. Direct numerical simulations for the turbulent mixing of the supersaturation field, which is assumed to be a passive scalar field in a homogeneous, isotropic  turbulent flow, were performed. For simplicity, we neglected the two-way coupling of the droplets, since the particle Reynolds numbers still remained smaller than one. Furthermore, we kept the flow in a statistically stationary regime to avoid complex transients which have been analysed for example in ref. \cite{Goetzfried2017}, and which would always be present in a real cloud. The strongly simplified model opened doors for a connection of the supersaturation and droplet statistics to the local strain statistics along Lagrangian trajectories which is quantified by the FTLEs. Our study is motivated by the aggregation model of turbulent mixing of scalar fields \cite{Villermaux2003,duplat2008mixing}.  

In our model, Eulerian and Lagrangian approaches are combined. The Lagrangian frame is incorporated via the cloud droplets (and thus the liquid water content) which are defined as Lagrangian tracers. Their size depends strongly on the fluctuations of the supersaturation field while they are advected by the turbulent flow. It was shown that the droplet size broadening strongly depends on the Reynolds number. A higher turbulence level triggers larger fluctuations of the supersaturation field and wider distributions of droplet radii. It was also shown that inside the cloud bulk, distribution of droplet radii are Gaussian without extended tails as found in entrainment studies \cite{Goetzfried2017}. the same holds for the distribution of the supersaturation. 

Different mixing regimes have been classified by the Damköhler number. In our DNS, we have an inhomogeneous mixing for the supersaturation field, ${\rm Da}_s>1$. It implies that regions are saturated much faster than they are mixed by the flow. For droplet evaporation process, a homogeneous mixing regime is obtained for the present parameter settings; in the bulk, droplets experience a rapid change of sub- and supersaturated regions along their pathways which cause ${\rm Da}_d < 1$. See again Fig. \ref{fig:Da_space}.

Using Lagrangian approach allowed us to obtain stretching and compression by calculating FTLE for our Lagrangian tracers. The interdependence of stretching regions defined by highest absolute values of the third FTLE and regions with highest supersaturation are obtained from a joint distribution of $\lambda_3$ and $s$. We have seen that the aggregation model of scalar filaments can be applied in principle, similar to \cite{Villermaux2003,gotzfried2019comparison} even though the Schmidt numbers is Sc $\sim 1$. Multiple foldings of scalar filaments are however unlikely; the diffusion times are short and a viscous-convective range of scalar mixing is very small. The statistics of droplet size distribution in the bulk of the cloud remains Gaussian. Furthermore, the small Reynolds numbers of the present cases generated Gaussian distributions of $\lambda_3$. At higher Re, intermittency in the inertial cascade range will generate asymmetric distributions in correspondance with large-deviation theory. These studies are currently taken towards the cloud boundary where the entrainment of clear air into cloudy air proceeds and the droplet size distributions develop tails. The results with expected deviations from a Gaussian statistics will be reported elsewhere.     

\acknowledgments
This work was supported by ITN CoPerMix. The project has received funding from the European Union’s Horizon 2020 research and innovation program under the Marie Skłodowska-Curie grant agreement N°956457. Supercomputing time has been provided at the University Computer Center (UniRZ) of the TU Ilmenau. The authors also gratefully acknowledge the Gauss Centre for Supercomputing e.V. (https://www.gauss-centre.eu) for funding this project by providing computing time on the GCS Supercomputer SuperMUC-NG at Leibniz Supercomputing Centre (https://www.lrz.de). We thank Priyanka Maity and Dmitry Krasnov for discussions.

\section{Appendix: Balance equation for the supersaturation field}
In the following we describe in brief the derivation of the equation of motion of the supersaturation field $s({\bm x},t)$ following the works of \citet{sardina2015continuous} and \citet{fries2021key}. Starting point are the balance equations of the temperature fluctuations $\theta({\bm x},t)=T({\bm x},t)-\langle T(z)\rangle$ and vapor mixing ratio fields $q_v({\bm x},t)$. They are given by
\begin{subequations}
    \begin{align}
        \label{eq:vapour_eq}
        \frac{Dq_v}{Dt}&=D_v \nabla^2q_v -C_d,\\
        \label{eq:temperature_eq}
     \frac{D\theta}{Dt} &=  \kappa \nabla^2 \theta - \frac{g}{c_{p}}u_z + \frac{L}{c_{p}}C_{d},
    \end{align}
\end{subequations}
where $D_v$ the vapor diffusivity, $\kappa$ the temperature diffusivity, and  $C_{d}$ the condensation rate. An additional term which contains the dry adiabatic lapse rate $g/c_p$ is contained in the equation of the temperature fluctuations. In clouds, the lapse rate is somewhat smaller than the dry one, $0.01\frac{K}{m}$; therefore even for the largest computation box from Tab. \ref{table:parameter_study} with size $L=0.512 m$ the mean temperature variation does not exceed $0.005 K$. Thus the prefactor in the updraft term of the equation for the supersaturation field will have a small prefactor $A_1$. We are not including a mean updraft of the volume as a whole with a velocity $\langle u_z\rangle$. The condensation rate is given by
\begin{equation}
\label{eq:cond_rate}
C_{d}({\bm x},t) =  \frac{1}{m_{a}}\frac{dm_{L}}{dt}=\frac{4 \pi \rho_{L}}{\rho V_{a}} \sum_{i=1}^N r_{i}^2 \frac{dr_{i}}{dt}
       =\frac{4 \pi \rho_{L} K'}{\rho V_{a}} \sum_{i=1}^N  r_i(t) s({\bm x}_{i},t)\,.
\end{equation}
We will now reduce the two equations to one balance equation, since the supersaturation field summarizes effectively the effects of latent heat release and condensation/evaporation in $q_{v}({\bm x},t)$ and $\theta({\bm x},t)$. We start with the material derivative of $s({\bm x},t)$ which is given by
    \begin{equation}
        \label{eq:supersaturation_derivative}
        \frac{Ds}{Dt} = \frac{D}{Dt}\left(\frac{q_{v}({\bm x},t)}{q_{vs}({\bm x},t)} - 1 \right) = \frac{1}{q_{vs}}\frac{Dq_{v}}{Dt} - \frac{q_{v}}{q_{vs}^2}\frac{Dq_{vs}}{Dt}.
    \end{equation}
First, the material derivative of the saturation mixing ratio (of vapor) $q_{vs}$, using the ideal gas law, gives
\begin{equation}
\label{eq:big_eq}
\frac{Dq_{vs}}{Dt}=\frac{D}{Dt}\left(\frac{\rho_{vs}}{\rho}\right)= \frac{D}{Dt}\left(\frac{\epsilon e_{s}}{p}\right) =\frac{\epsilon}{p} \frac{De_{s}}{Dt} - \frac{\epsilon e_{s}}{p^2} \frac{Dp}{Dt} =
 \frac{\epsilon}{p} \frac{de_{s}}{d\theta} \frac{D\theta}{Dt} - \frac{\epsilon e_{s}}{p^2} \frac{Dp}{Dt},
\end{equation}
where $\varepsilon= R'/R_{v}$. The total time derivative of the pressure field can be obtained via the hydrostatic equilibrium \cite{pruppacher1998microphysics}
\begin{equation}
    \frac{Dp}{Dt}=\frac{dp}{dz}\frac{dz}{dt}=-u_z g \rho = - \frac{u_z g p}{R_a\theta},
    \label{eq:dpdt}
\end{equation}
where $R_a$ is gas constant for dry air. By using the Clausius-Clapeyron equation \eqref{eq:dpdt} and (\ref{eq:temperature_eq}),  eq. (\ref{eq:big_eq}) is transformed to
    \begin{equation}
        \label{eq:saturated_vapor_eq}
        \frac{Dq_{vs}}{Dt} = \frac{\epsilon}{p} \frac{L e_{s}}{R_{v} \theta^2} \left(\kappa \nabla^2 \theta - \frac{g}{c_{p}}u_z + \frac{L}{c_{p}}C_{d}  - \frac{\epsilon e_{s}}{p^2}\frac{u_z g p}{R_a\theta}\right).
    \end{equation}
Substituting (\ref{eq:saturated_vapor_eq}) and (\ref{eq:vapour_eq}) into (\ref{eq:supersaturation_derivative}) results to
\begin{equation*}
\frac{Ds}{Dt} = \frac{1}{q_{vs}}\left(D_q \nabla^2 q_v - C_{d} \right) - \frac{q_{v}}{q_{vs}^2}\frac{\epsilon}{p} \frac{L e_{s}}{R_{v} \theta^2} \left(\kappa \nabla^2 \theta 
-\frac{g}{c_{p}}u_z + \frac{L}{c_{p}}C_{d}- \frac{\epsilon e_{s}}{p^2}\frac{u_z g p}{R_a\theta} \right)\,.
\end{equation*}
Thus follows
\begin{equation*}
\frac{Ds}{Dt} = \frac{1}{q_{vs}}\left(D_q \nabla^2 q_v - C_{d} \right) - \frac{s+1}{q_{vs}}\frac{\epsilon}{p} \frac{L e_{s}}{R_{v} \theta^2} \left(\kappa \nabla^2 \theta -\frac{g}{c_{p}}u_z + \frac{L}{c_{p}}C_{d}- \frac{\epsilon e_{s}}{p^2}\frac{u_z g p}{R_a\theta} \right).
\end{equation*}
For clouds supersaturation usually does not exceed $1\%$--$2\%$. Thus $s\ll 1$ and consequently $s+1 \approx 1$. Together with the definition of the saturation mixing ratio, one gets
\begin{equation}
\label{eq:s_pre_eq}
\frac{Ds}{Dt} = \frac{p D_q}{\epsilon e_{s}} \nabla^2 q_v - \frac{p}{\epsilon e_{s}}\frac{L e_{s} \kappa}{R_{v} \theta^2} \nabla^2 \theta-
C_{d}\left(\frac{p}{\epsilon e_{s}} + \frac{p}{\epsilon e_{s}} \frac{L^2 \epsilon e_{s}}{p R_{v} \theta^2 c_{p}}\right)+  \frac{L g}{R_{v} c_{p} \theta^2}u_z\,.
\end{equation}
In the present model, we will approximate diffusion of the supersaturation field as a linear combination of the diffusion processes of temperature and vapour mixing ratio,
\begin{equation}
\label{eq:s_diff}
D_{s} \nabla^2 s = \frac{p D_q}{\epsilon e_{s}} \nabla^2 q_v - \frac{p}{\epsilon}\frac{L \kappa}{R_{v} \theta^2} \nabla^2 \theta.
\end{equation}
The diffusion coefficient of supersaturation is assumed to be equal to the diffusion coefficient of water vapour, $D_{s} \approx D_q$. Thus the Schmidt number of the scalar supersaturation field will also be $Sc\approx 0.7$. The coefficient which is connected to the condensation rate can then be transformed in the following way
\begin{equation}
\label{eq:coeff_at_cd}
\frac{p}{\epsilon e_{s}} + \frac{p}{\epsilon e_{s}} \frac{L^2 \epsilon e_{s}}{p R_{v} \theta^2 c_{p}} = \frac{\rho R' \theta}{\epsilon e_{s}} + \frac{L^2 \epsilon  \rho \theta}{p \theta^2 c_{p}} = \rho \left(\frac{R' \theta}{\epsilon e_{s}} + \frac{L^2 \epsilon}{p \theta c_{p}} \right).
\end{equation}
Substituting (\ref{eq:s_diff}), (\ref{eq:coeff_at_cd}) and (\ref{eq:cond_rate}) to (\ref{eq:s_pre_eq}), we get the final equation for the supersaturation field which is given by
\begin{equation} 
\frac{Ds}{Dt} = D_{s} \nabla^2 s +\frac{L g}{R_{v} c_{p} \theta^2}u_z 
- \left(\frac{R' \theta}{\epsilon e_{s}(\theta)} + \frac{L^2 \epsilon}{p \theta c_{p}} \right)\frac{4 \pi \rho_{L} K}{V_{a}} \sum_{i=1}^N  r_i(t) s(\textbf{x}_{i}, t).
\label{eq:supersaturation}
\end{equation}
The following coefficients can then be defined,
\begin{subequations}
\label{eq:therm_coeff}
\begin{align}
A_1 &= \frac{L g}{R_{v} c_{p} \theta^2},\\
A_2 &= \frac{R' \theta}{\epsilon e_{s}} + \frac{L^2 \epsilon}{p \theta c_{p}},\\
K' &= \left( \frac{L \rho_{L}}{K\theta}\left(\frac{L}{R_{v} \theta} - 1\right) + \frac{\rho_{L}R_{v}\theta}{D e_{s}(\theta)} \right)^{-1}.
\end{align}
\end{subequations}
Thus the final balance equation of the supersaturation field is
\begin{equation}
\frac{Ds}{Dt} = D_{s} \nabla^2 s + A_1 u_z - A_2 \frac{4 \pi \rho_{L} K'}{V_{a}} \sum_{i=1}^N  r_i(t) s(t,\textbf{x}_{i}).
\end{equation}
This equation couples Eulerian and Lagrangian dynamics and is used in the main text.
\nocite{*}
\bibliography{main}
\end{document}